# High Capacity, Secure (n, n/8) Multi Secret Image Sharing Scheme with Security Key

Karthik R, Tejas K, Swathi C, Ashok K, Rajesh Kumar M, *Senior Member, IEEE*
Department of Electronics and Communication Engineering
VIT University
Vellore, India 632014
Email: tejastk.reddy@gmail.com, mrajeshkumar@vit.ac.in

*Abstract*— The rising need of secret image sharing with high security has led to much advancement in lucrative exchange of important images which contain vital and confidential information. Multi secret image sharing system (MSIS) is an efficient and robust method for transmitting one or more secret images securely. In recent research, *n* secret images are encrypted into *n* or *n+ 1* shared image's and stored in different database servers. Decoder has to receive all n or n+1 encrypted images to reproduce the secret image. One can recover partial secret information from *n-1* or fewer shared images, which poses risk for the confidential information encrypted. In this proposed paper we developed a novel algorithm to increase the sharing capacity by using (*n, n/8*) multi secret sharing scheme with increased security by generating unique security key. An unrevealed comparison image is used to produce shares which makes the secret image invulnerable to the hackers.

## I. INTRODUCTION

Now a day's growth of multimedia and internet has made digital communication one of the prime means of communication in all sectors of the society from corporates to citizens. It plays an important role especially in the fields of secret intelligence services, patient's medical information transfer, army and scientific laboratories etc. As a result, security of digital information against unauthorized access has become a prime objective. In order to tackle this problem, various different technologies evolved in the recent times such as steganography, encryption, multi-secret image sharing scheme, cryptography, perceptual hashing and digital watermarking. Cryptography is the process of converting plain text to cipher text by the sender with the use of an encryption key and other side receiver decrypts cipher text to plain text with the help of the security key. Encryption is the process of slicing a secret image into *n* shares and the process of recovery of original image by overlapping all the *n* shares is called decryption.

Multi secret image sharing system is a part of visual cryptography which was introduced by Naor et al [1], where a secret image is bit sliced and encrypted into various multiple shares which do not disclose any information independently.

Chin-Chen et al [2] proposed the concept of secret sharing by hiding two spatial-domain images. This schemes uses the two-out-of-two visual secret sharing technique to generate two shares for hiding a secret two-to-one image and also proposed an embedded scheme to put two secret shares into two grey-level cover images. Chai-Chun et al [3] proposed a novel secret image sharing scheme based on the simple LSB substitution and an optimal pixel adjustment process. Yi Ching Zeng et al [4] proposed scheme depicted multi scale image sharing scheme to hide multiple images to two meaningful sharing images and improve capacity of hidden image, which combines visual cryptography with data hiding technique. Hongliang et al [5] proposed that all the generated share image is meaningful image and same size as the secret image. Gayathri et al[] proposed scheme that can share one. Digital secret image using *n-1* selected share this method needs only one share image to share one secret image it is generated using the secret image and natural image. Lakshmi et al [6] proposed that substitution of least significant bits which results in high embedding capacity human eyes are sensitive to change in smooth region. Hence LSB substitution scheme is not a good choice for embedding data.it can hide two secret image in to a single cover image and also satisfied the reversibility property of steganography to recover the secret and cover image without any loss of information. Maroti et al [7] proposed a method which has more efficient and secure *(n, n)* MSIS scheme so that less than *n* shared images does not reveal any information of secret image using XOR and modular arithmetic's and randomness is increased by reverse bit function.

Here, the sharing capacity is '1', i.e. all *n* encrypted shares need to be sent to decoder to retain the secret image. Transferring all *n* shares is a cumbersome task which also involves the risk of losing shares or being hacked is high. Thus in order to increase the sharing capacity, we propose a novel algorithm where the sharing capacity is increased by eight times and also increasing the security by using an encryption security key. Additional high security is ensured by not sending the comparison image used for encryption of shares to the decoder, rather the image name is conveyed to decoder in a highly secure medium of communication like person to person communication. This image can be downloaded from the internet by the decoder and can be used to decode this multi-secret image share. This algorithm can be best suited for video sharing where every eight frames can be sent as one secret image.

Single image sharing algorithms produce *n* shares from one secret image. To recover this secret image by the decoder, all *n* shares are required. This algorithm poses inconvenience to the user to share many images. In order to solve this issue, multi- secret image sharing (MSIS) algorithms were developed. An (*n, n*) MSIS algorithm can send one share per secret image, which also poses inconvenience to many images or video frames as secret data. This issue can be relatively solved by the proposed algorithm which allows the user to transmit eight image shares at a time.

## II. PROPOSED METHOD

### 1. Security key

The use of secret key in producing shares from image can add an additional security layer as compared to reverse bits algorithm as used in the earlier works. This secret key is produced from the unrevealed comparison image through algorithm 1. This key is later used to produce shares from the secret images. The decoder has to apply algorithm 1 on the comparison image to produce the key, for further use in recovering all the secret images.

---

### Algorithm 1

**Input**: Comparison image.
**Output**: Security key.

Step 1: Extract pixel by pixel from first row i.e. (0, $A_i$) where $A_i = \{0,1,2,3....n\}$.

Step 2: Divide the extracted pixel value by 8, and record the remainder.

Step 3: Create a linear array of 8 element size.

Step 4: Check if the remainder value is existing in the array,
If exists = discard the value, increment 'i' and repeat from step 1.
If not exists then place the remainder value in the array from left to right and repeat from step 1.

Step 5: Once all the eight elements in the array are filled, exit from the loop.

---

When a pixel value is divided by 8, the remainder will be a value from 0 to 7. Thus, 8 distinct numbers, ranging from 0 to 7 are placed in the linear array to form the security key, which we later use for further production of shares.

### 2. Encryption

We now produce *n/8* meaningless shares from n input images, which need to be transferred as secret images. These *n/8* shares are produced in such a way that there is no partial information leaked when *n/8 – 1* images are embedded. This can be achieved by applying algorithm 2 on the images to be hidden.

---

### Algorithm 2

**Input**: *n* images that need to be transferred secretly and the comparison image.
**Output**: *n/8* number of secret images.

Step 1: The comparison image ($I_1$) is now bit sliced according to the order defined by security key.
If security key is [70452316], the MSB, LSB, $4^{th}$, $5^{th}$, $2^{nd}$, $3^{rd}$, $1^{st}$ and $6^{th}$ layer are extracted in same order mentioned.

Step 2: Each layer extracted is then converted into a linear array and are placed side by side in a linear matrix $L_1$.

Step 3: Input the image to hide 1 = $I_2$.

Step 4: Extract each pixel from $I_2$, convert it into binary number and add side by side in a linear matrix $L_2$.

Step 5: Size of ($L_1$) = size of $L_2$.
If extra elements are found in $L_1$, discard them.
If less number of elements is found in $L_1$ than $L_2$, pad zeros to cover the shortage elements.

Step 6: XOR corresponding elements of $L_1$ and $L_2$ and store result in $L_3$.
$L_1 \oplus L_2 = L_3$.

Step 7: Considering each element in linear matrix $L_3$ as a pixel, create a square matrix or a binary image with equal number or rows and columns = $I_3$.

Step 8: Consider the images to be hidden (of same size of $I_2$), for each image repeat from step 2 to step 7.

Step 9: Now for every eight secret images to hide, we have eight binary images formed in step 7.

Extract the corresponding pixel values in all these 8 images formed in step 7, convert the binary digits to decimal value, and place this decimal in a new image which is the same size of ($I_3$).

---

In the end of step 9 of algorithm 2. We get a grey scale image $I_4$ which acts as a secret image share containing information of eight other images. If the sender has to send *n* images to receiver, where *n* is not a multiple of eight, null images can be padded to get the value of n to a multiple of eight.

This image $I_4$ alone can be sent to the decoder to decode all the eight hidden images in it. In order to have an additional security layer we do not recommend to send the comparison image to the receiver in the same channel where the secret image was sent, instead we could tell the receiver by person, the name of the image or the link from where he could download that from internet which adds more security from the hackers.

### 3. Decryption

Once the decoder has the knowledge of comparison image, algorithm 1 is applied onto the image to get the security key. In order to retrieve all eight secret images from every image share, we apply the reverse of algorithm 2, as explained in detail by algorithm 3.

---

**Algorithm 3.**

**Input**: *n/8* meaningless secret image shares.
**Output**: n retrieved secret images.

Step 1: Input a secret image share=$I_4$

Step 2: Create 8 dummy binary images of the same size of $I_4$.

Step 3: Extract each pixel from $I_4$, convert it into binary form. The eight binary digits are then replaced in 8 images in corresponding locations of the extracted pixel.

Step 4: Consider each image individually, and convert it into a linear matrix $L_5$.

Step 5: Matrix $L_1$ is calculated from comparison image similar to step 1&2 in algorithm.2.

Step 6: Now $L_5 \oplus L_1 = L_6$, where $L_6$ is a linear matrix.

Step 7: $L_6$ linear matrix is then converted back into the secret image $I_2$.

Step 8: Step 4 to step 8 is repeated for all eight images formed in step 3 to get all images retrieved from the meaningless share.

---

We obtain eight hidden secret shares and their corresponding images for every transmitted share from the sender.

## III. EXPERIMENTAL RESULTS

In this section, the 'security' and 'capacity' of the proposed algorithm is evaluated. These evaluated results are compared with earlier works for better understanding. The proposed algorithm was executed in MATLAB (R2016a) software.

(n, n/8) Multi-secret image sharing scheme can share n secret images among n/8 shared images. No partial information is gained if two or more shares are embedded since every share has its individual properties, which makes the algorithm efficient towards maintaining confidentiality. This proposed scheme also has advantage towards sharing capacity.

$$sharing\ capacity = \frac{no.\,of\ secret\ images}{no.\,of\ shared\ images} \quad (1)$$

$$sharing\ capacity = \frac{n}{\frac{n}{8}} \quad (2)$$

$$sharing\ capacity = 8 * \frac{n}{n} \quad (3)$$

This proves that the sharing capacity of the proposed scheme is eight times the *(n, n)* MSIS scheme i.e. this algorithm is capable of transmitting eight secret images as one meaningless share rather than transmitting one secret image as a share. This makes the algorithm more compatible in sending bulk images or video frames.

The proposed *(n, n/8)*-MSIS scheme algorithm is well suited for both grayscale and colored images. For better experimental analysis grayscale images are used. Fig.1. illustrates the proposed *(n, n/8)*-MSIS algorithm, where *n=8*. The input eight images are taken from AT&T official face data base as shown in Fig.1.(a) which is converted into a secret image share by processing through algorithm 2. Here, 'Lena.jpg' is used as the unrevealed comparison image which is used to produce the meaningless shares. The individual shares Fig.1.(c) as well as the final share Fig.1.(d) both do not reveal any partial information individually, which adds to the confidentiality. The final secret share produced is now transmitted to the decoder and the comparison image is revealed to the decoder in a safe communication medium, which is used by the decoder to produce the secret key by using algorithm 1. This is further processed through algorithm 3 to recover all the meaningless shares from the transmitted share image as shown in Fig.1.(e) from which the original secret images are restored as shown in Fig.1.(g).

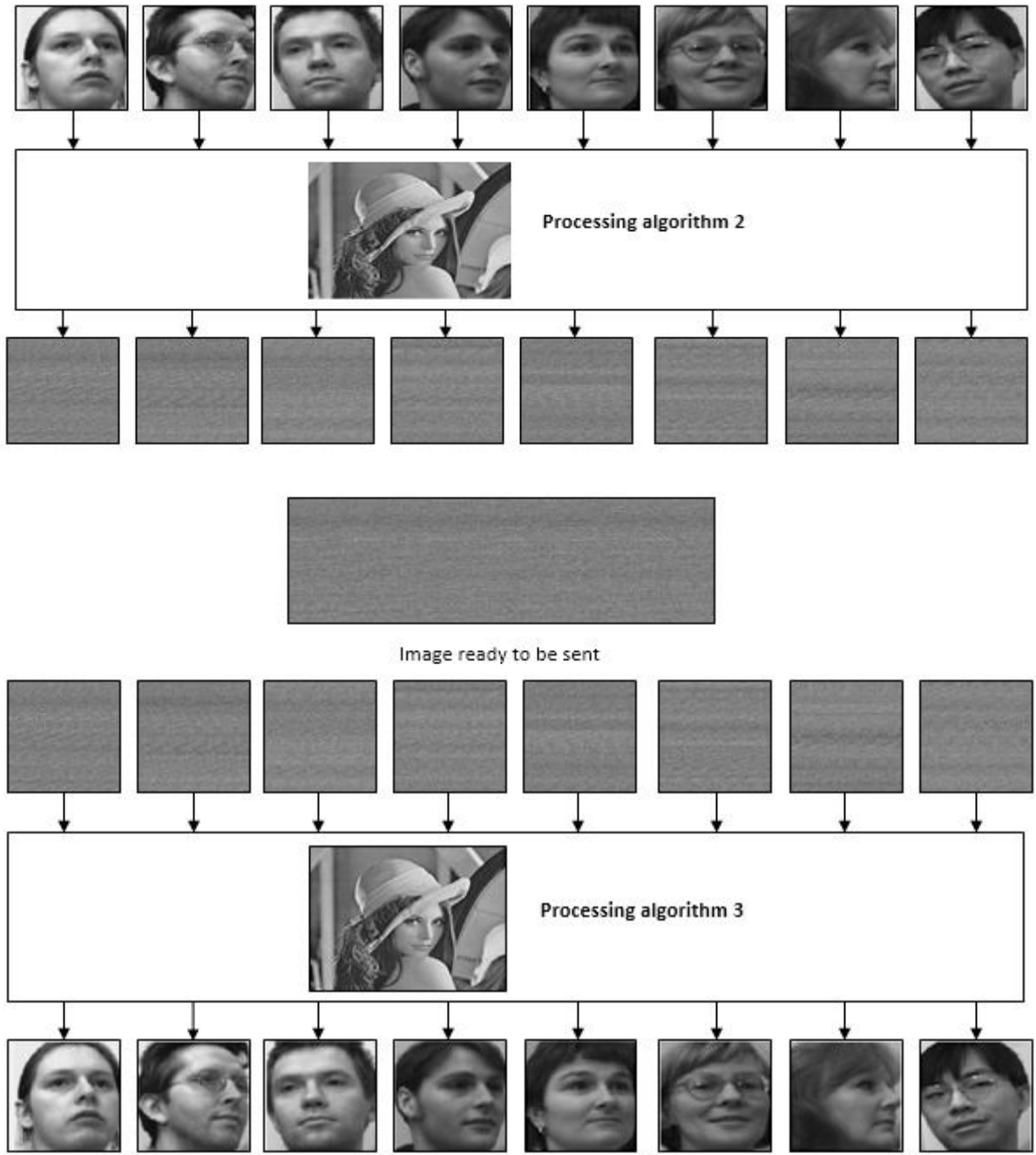

Fig 1.Flowchart of the proposed algorithm. Rows (a),(b),(c) and (d) illustrates the encryption algorithm whereas (d),(e),(f) and (g) illustrate the decryption algorithm. (a) shows the secret images to be sent. (c) shows the individual image shares produced. (d) shows the embedded image ready to be sent. (e) shows the shared obtained from (d). (g) shows the recovered secret images by the decoder.

Advantages:
1. Sharing capacity is increased by 8 times due to which it is easier to transfer the shares while transferring large datasets of images.
2. Since the comparison image need not to be not sent directly, the secret image share remains invulnerable to the hackers.
3. The use of security key increases the randomness of the shares as compared to reverse bits algorithm.
4. No partial information disclosed by secret image share formed in algorithm 2.

Table I

COMPARISON OF EXISTING MSIS AND PROPOSED *(n, n/8)* SCHEMES FOR GRAYSCALE IMAGES WITH *n =8*

| Parameters | Chen and Wu[9] | Chen and Wu[10] | Yang[8] | Maroti [7] | Proposed method |
|---|---|---|---|---|---|
| Secret image | n | n | n | n | n |
| Shared image | n+1 | n | n | n | n/8 |
| Sharing capacity | n/n+1 | n/n | n/n | n/n | 8(n/n) |
| Pixel expansion | no | no | no | no | no |
| recovery | lossless | lossless | lossless | lossless | lossless |
| Reveals secret | partial | partial | partial | no | No |
| Sharing type | rectangle | rectangle | rectangle | rectangle | square |
| Color depth | Grey | Grey | Grey or color | Grey or color | Grey or color |
| Recovery strategy | XOR | XOR | XOR | XOR | XOR with unrevealed comparison image |
| Increasing randomness of shares | – | – | – | Reverse bits | Security key |

The qualitative experimental results obtained in the proposed algorithm are illustrated in Table 1. These qualitative parameters and experimental key factors are discussed and compared with earlier state-of-art works. It is clearly observed that proposed algorithm proves its efficiency in parameters like sharing capacity, recovery strategy and procedures used for increasing randomness of shares.

Table II

Matching between secret and recovered images

| Input image | Recovered images | SSIM | PSNR | RMSE |
|---|---|---|---|---|
| $F_1$ | $F_1'$ | 1.00 | Inf | 0 |
| $F_2$ | $F_2'$ | 1.00 | Inf | 0 |
| $F_3$ | $F_3'$ | 1.00 | Inf | 0 |
| $F_4$ | $F_4'$ | 1.00 | Inf | 0 |
| $F_5$ | $F_5'$ | 1.00 | Inf | 0 |
| $F_6$ | $F_6'$ | 1.00 | Inf | 0 |
| $F_7$ | $F_7'$ | 1.00 | Inf | 0 |
| $F_8$ | $F_8'$ | 1.00 | Inf | 0 |

$F_x$- Input Image , $F_x'$-Recovered Image  x∈ [1,8]

The proficiency of qualitative results obtained from the proposed algorithm are illustrated in Table.2. Here, the efficiency of recovering the encrypted secret image is verified. This verification is done by

1. Structure similarity check.
2. PSNR
3. RMSE

From Table.2. we can see that recovering efficiency is 100% for all the eight images embedded in the share.

## IV. CONCLUSION

Sharing many images using MSIS from sender to receiver was a cumbersome task and involved the risk of losing shares or being hacked in the channel. The proposed algorithm paves an efficient way to transmit n secret images in n/8 shares. This algorithm is best suited to share multiple frames in a video. Usage of security key concept acts as an additional security layer for being hacked. The secret image remains invulnerable for the hackers without the information about the comparison image. The experiment results prove the efficiency of the proposed algorithm over the existing state-of-art methods.